\documentstyle[amsmath,emulateapj]{article}

\lefthead{Romeo et al.} \righthead{Millimetric Ground Based
Observation of CMBR Anisotropy at $\delta=+28^{o}.$}

\begin{document}

\title{Millimetric Ground Based Observation of\\
CMBR Anisotropy at $\delta =+28^{\lowercase{o}}$}
\author{G. Romeo\altaffilmark{1},S. Ali\altaffilmark{1},
 B. Femen\'{\i}a,\altaffilmark{2}
M. Limon,\altaffilmark{3}  \\
%EndAName
L. Piccirillo\altaffilmark{1,4,5}, R. Rebolo\altaffilmark{2}, R. Schaefer\altaffilmark{6}}

\altaffiltext{1}{Bartol Research Institute, University of Delaware, Newark,
DE, 19716}
\altaffiltext{2}{Instituto de Astrof\'{\i}sica de Canarias, 38200 La
Laguna, Spain}
\altaffiltext{3}{Princeton University, Physics Department,
Jadwin Hall, Princeton, NJ 08544 } %
\altaffiltext{4}{University of Wisconsin, Madison, WI, 53706} %
\altaffiltext{5}{current address: Department of Physics and Astronomy, Cardiff
University, Cardiff CF24 3YB, UK}
\altaffiltext{6}{Raytheon ITSS, NASA Code 664, Goddard Sp. Fl.
Ctr., Greenbelt, MD 20771}

\begin{abstract}
Results from the third campaign of a ground-based multi-band
observation of the millimeter emission of the sky from Tenerife
(Canary Islands) are presented. The instrument consists of a 0.45
meter diameter off-axis telescope equipped with a 4-band
multi-mode $^3He$ cooled photometer working at 1.1, 1.3, 2.1
and 3.1{\notetoeditor{These are effective wavelengths}} mm wavelengths.
The beam is well approximated by a Gaussian with 1$%
^{o}$.35 Full Width Half Maximum (FWHM) at all wavelengths. The wide
wavelength coverage of our instrument allows us to characterize and reduce
both the atmospheric and galactic contamination in our data. The CMBR data
is analyzed in 6 multipole bands whose centers span the range $\ell=39$ to
$\ell =134$ at the two longest wavelengths (2.1 and 3.1 mm). A likelihood analysis
indicates that we have detected fluctuations in all bands at the two
wavelengths. We have evidence of a rise in the angular power spectrum from low $\ell $ to
high $\ell $. Our measured spectrum is consistent with current popular theories of
large scale structure formation, COBE, and other recent balloon-borne
experiments with similar wavelength coverage.

\end{abstract}


\section{Introduction}

Precise measurements of the angular power spectrum of the Cosmic Microwave
Background Radiation (CMBR) provide strong constraints on cosmological parameters.
After the era of the discovery of the anisotropy by COBE (\cite{Smoot92} and
\cite{Bennet96}) we are now in the age of spatial spectroscopy of the primordial
fluctuations. Many experiments \notetoeditor{see $http://www.sns.ias.edu/%
\sim max/cmb/experiments.html$ and (\cite{Page97})} have reported detections of
anisotropy at different angular scales with evidence of excess of power at
angular scales of about $\ell \sim 200$.

We designed an experiment sensitive to a broad angular range
($\ell$ = 39 to 134 ). Our lowest $\ell$-band $39^{+38}_{-24}$
partially overlaps with COBE's highest $\ell$-band. The angular range also overlaps
with COBE at our longest wavelength (3.1 mm). The partial overlap
of the parameters of our experiment and COBE allows us to perform
independent consistency checks.

In this letter, we present the data from our third observing campaign of
mm-wave CMBR anisotropy from Tenerife (Canary Islands). The first campaign
in 1993 was intended to test the instrumental set-up as well as characterize
the site. The second campaign in 1994 (see {\cite{Piccirillo97}} and {\cite
{fem97}}) measured the anisotropy at two $\ell $ bands. In the third
campaign we extended the angular scale coverage from 2 to 6 bands.
Observations were carried out from May to July, 1996.

\section{Instrument}

The 1996 Tenerife measurements were performed with the same basic
telescope of the 1994 campaign. However, the beam size has been
reduced to $1^o.35$ Full Width Half Maximum (FWHM). For details on
the instrument see (\cite {fem98}) (\cite{nicho96}),
(\cite{ali98}), and (\cite{romeo2000}). Here we briefly summarize its main
characteristics. The sky radiation is collected by the primary
mirror (0.45 meter diameter). The primary is sinusoidally chopped
at 2 Hz with a peak-to-peak amplitude of $5^o.7$ in the sky. The
collected radiation is then focused into the photometer by means
of a fixed off-axis secondary mirror (0.28 meter diameter).
The photometer consists of a 4-band He-3 bolometric
system operating at 1.1, 1.3, 2.1 and 3.1 mm wavelengths.
The voltage signal from each bolometer is sampled
synchronously with the sinusoidal motion of the primary mirror. The
sampling rate is 128 samples/channel/second corresponding to 64
samples/channel per sinusoidal cycle of the mirror. In order to
minimize systematic effects we surrounded the optics with two
levels of radiation shields.

\section{Calibration}

The primary calibration of the instrument was performed by
observing the mm-wave emission of the Moon. Several raster scans
of the Moon have been used together with a computer simulation to
fit the following parameters: beam size $(1^{o}.35)$, beam throw
$(5^{o}.7)$, the calibration constants in Volts/Kelvin for each
instrumental band and each demodulation, the azimuth and elevation
tracking and pointing accuracy.  A model for the Moon emission
temperatures was provided by ({\cite{Gulkis}}). A correction for
atmospheric attenuation in our bands was estimated from skydips as
well as from atmospheric models. The dominant calibration
uncertainty comes from the
error in the Moon temperatures $(\pm 5\%)$ and the atmospheric attenuation $%
(\pm 5\%)$. We add these two errors linearly to yield a total calibration
uncertainty of $\pm 10\%$.

\section{Observations}

Two different observing strategies have been used: the "Dec40" and the
"zenith" modes. The Dec40 mode consists
of observations in drift scan at a fixed elevation corresponding to $\delta
=40 ^{o}$. This declination has been extensively studied at lower frequencies
by the Tenerife collaboration (see \cite{guti99}) as well as from our
instrument in 1994. For the results of our 1996 campaign at declination 40$^{o}$ see the
companion paper (\cite{ali98}). The data discussed in this Letter are
collected during the ''zenith'' mode:the telescope is fixed to observe the
zenith and the sinusoidal chopping throws the beam along the North-South
direction. The East-West sky rotation provides the sampling of a strip of the
sky defined by $0h<RA\leq 24h$. This observing strategy is similar to the
Saskatoon experiment (see \cite{netterfield}), although we observe
at a different declination.

Data were collected from May through June 1996 when atmospheric fluctuations are at
their minimum. Any further contamination was minimized by
using only the data collected at night, when the Sun was well below
the horizon. About 241 hours of data (from 23 nights of
observation), corresponding to about 1000 square degrees of sky
observed at 4 different wavelengths were collected by using the
zenith-mode.

Our zenith observing strategy has the advantage that the telescope does not
move during the CMBR observations. This virtually eliminates any changes in
the sidelobe pick-up that can contaminate the data due to the motion of
the telescope.

\section{Data Reduction and Analysis}
Data are demodulated using the following set of orthogonal functions
\begin{equation}
L_{n}(\omega t)=\cos (n{\omega }t)+i\sin (n{\omega }t)={e^{(in\omega
t)}}\label{eqn:1}
\end{equation}
where $\omega$ is the chopping frequency. The output of the $n-th$ synthesized beam is:
\begin{equation}
a_{n}+{\itshape i}b_{n}={\frac{1}{2\pi }}{\sum_{k=1}^{N}L_{n}\biggl({%
\frac{2\pi k}{N}}\biggr )S_{k}}
\end{equation}
where $S_{k}$ are the discrete samples of the bolometer voltages, N=64 is
the number of samples per sinusoidal cycle and n=1...6 are the 6
demodulation vectors indices, which can be used to reconstruct a
2-dimensional map of anisotropy (\cite{romeo99}).
Our choice of orthogonal functions allows us to
express the demodulated output according to:
\begin{equation}
\begin{split}
a_{n}& ={\frac{1}{2\pi }}\sum_{k=1}^{N}S_{k}\cos \biggl( {\frac{2\pi nk}{N}}%
+\psi _{n}\biggr ) \\
b_{n}& ={\frac{1}{2\pi }}\sum_{k=1}^{N}S_{k}\sin \biggl( {\frac{2\pi nk}{N}}%
+\psi _{n}\biggr )
\end{split}
\end{equation}
where the $\psi _{n}=\tan ^{-1}(b_{n}/a_{n})$ are phases which account for the
instrumental delays between the time ordered data $S_{k}$ and the position
of the beam in the sky. Observations of the Moon signal determine these
phases with high accuracy. Each night of observation produces 6 demodulated
files per bolometric channel.

The first step in our analysis is to subtract the atmospheric noise
from channels 1, 2, and 3 using the information in channel 4, which is
the most sensitive to atmospheric noise.  Our technique is essentially
the same as used in the analysis of our 1994 campaign data
(\cite{nicho96}, \cite{fem97}, and \cite{Piccirillo97}) which was already an
improved version of the technique used by \cite{Andreani}.  The
details of the procedure are to be found in \cite{romeo2000} and \cite{ali98}.
A more general theoretical discussion can be found in \cite{Melchiorri}.
The atmospheric subtraction can only be applied when the
atmospheric conditions are stable and there is a strong correlation
($>75\%$) between channel 4 and the other channels.   In the lower
harmonics, the power spectra of the atmospheric subtracted data show a
strong improvement in the signal to noise ratio.   For the higher
harmonics (n = 4, 5, and 6), we  did not need the atmospheric
subtraction because the high degree of spatial differencing breaks up
the atmospheric noise, whose power decreases sharply with scale.
(Atmospheric noise follows a Kolmogorov power spectrum which decreases
as the wavenumber to the -8/3 power. )
 A more in-depth analysis of the atmospheric noise will be discussed in
even greater detail in \cite{Ali2000} and \cite{ali98}.

Long term drifts in the data are then removed
by fitting a combination of sine and cosine functions(\cite{fem97} and
\cite{fem98}). The spatial
frequencies used correspond to angles of 90, 60, 30, 30, 30 30
degrees R. A. for demodulations 1-6 respectively, angular scales
where the window functions are negligible. Data are finally stacked
together and binned in RA intervals of about $1^{o}.4$.

The final data set is analyzed by performing a likelihood analysis of each
individual demodulation for the two longest wavelength bands (3.1 and 2.1
mm) corresponding to 6 $\ell $ ranges centered from $\ell $=39 to $\ell $=134.
In fig. {\ref{fig1}} we show
the first demodulation of the bands together with the demodulations
obtained by simulating our observing strategy on the DIRBE 240 $\mu m$ map . %
\placefigure{fig1} The feature visible in the DIRBE demodulations at
RA about 20h is the region corresponding to the crossing of the galactic
plane. We see that the same feature is visible in the 1.3 mm band and
somewhat less visible with increasing wavelength to 2.1 and 3.1 mm.

We can use the ratios of intensities I(DIRBE)/I(mm) to extrapolate the
galactic signal in the region of $12<RA<19$ where we analyze the data for
CMBR anisotropy. The result of the extrapolation is shown in fig. {\ref
{fig2}}. \placefigure{fig2} We see that, at high galactic latitude, in the
two longest wavelength bands (1 and 2) the galactic contamination
should be negligible while the band at 1.3 mm might be contaminated by residual
galactic dust emission and is not included in this analysis.

Computing the various window functions is the last step needed to
perform the likelihood analysis. The relative simplicity in the
observing strategy comes at the cost of some algebra in computing
the window functions for each individual demodulation at any
angular lag. Following (\cite{White94}), the expression for the
window function of each demodulation is:
\begin{equation}
\begin{split}
& W_{\ell}^n(\hat{k}_{i}\cdot \hat{k}_{j}) =\aleph^2 B_{{\it l}}^2 %
\Biggl \{ \biggl [Q_{\ell,0}^n (\theta_o) \biggr ]^2 + 2 \sum_{m=1}^{\ell}
{\frac{(\ell-m)!} {(\ell+m)!}} \times\\
&\quad\quad\quad \cos(m(\phi_i -\phi_j))
\Biggl [Q_{\ell,m}^n (\theta_o)
{\frac{\sin ( {\frac {m \Delta \phi} {2}} )} {\frac{m
\Delta \phi} {2}}} \Biggr ]^2 \Biggr\}
\end{split}
\end{equation}
where $n$ is the integer for the corresponding demodulation,
$\hat{k}_{i}$ and $\hat{k}_{j}$ are respectively the unit vectors
identifying right ascension bin $i$ and $j$, $\aleph$ is the
normalization determined in the same way as for Saskatoon [see
\cite{White94}],

\[
B_{\ell}(\sigma) = \exp \biggl [ - {\frac {1} {2}} {\ell}({\ell}+1) \sigma^2
\biggr ]
\]
is the beam profile function with $\sigma$=0.$^{o}$57, our
Gaussian beam width, and
\[
Q_{\ell,m}^n (\theta_o) = {\frac {\omega_c} {2 \pi}} \int_{-\pi/\omega_c}^{%
\pi/\omega_c} L_n(t) P_{\ell}^{{\it m}} [\cos(\theta(t))] dt \quad\quad \text{m
= 0...{$\ell$}}
\]
where $L_n(t)$ is the lock-in function in equation \ref{eqn:1}, $P_{\ell}^{{\it m}}
[\cos(\theta(t))] $ is the associated Legendre polynomials, $%
\Delta \phi$ is the binning size in Right Ascension, and finally
$\theta(t)=\theta_o+\alpha \sin(\omega t)$, with $\theta_o=28^o$,
and $\alpha=2^o.84$.

The window functions are estimated numerically.

\section{Statistical analysis}

The statistical analysis used to determine the amplitude of the fluctuation
in the CMB was performed by using the Likelihood defined as follows:
\begin{equation}
{\mathcal{L}} ={\frac{1}{(2\pi )^{N/2}|{\bf {M}}|^{1/2}}}\exp \biggl (-{\frac{1}{2}%
}t^{T}{\bf M^{-1}}t\biggr )
\end{equation}
where t is the vector containing the demodulated and stacked data.
We maximize the likelihood L for channels 1 and 2 both individually and
collectively.  Each channel has 72 1.4 degree bins of stacked data for
each harmonic, so the covariance matrix M is a 72 x 72 (144 x 144)
matrix for the individual (combined) likelihood, respectively.  The
combined channel analysis includes terms for the correlation
between channels.
   The covariance matrix is the sum of two matrices, signal $(S_{ij})$ and
noise $(N_{ij})$.  The signal is given by

\[
S_{ij}={\frac{1}{4\pi }}\sum_{\ell =1}^{\infty }(2{\ell }+1)C_{\ell }W_{\ell
}^{n}(\hat{n}_{i}\cdot \hat{n}_{j})
\]
where $W_{\ell }^{n}(\hat{n}_{1}\cdot \hat{n}_{2})$ is the window function,
and $C_{\ell}$ are the usual temperature angular correlation
coefficients. The noise covariance matrix is
\[
N_{ij}=\sigma _{i}\sigma _{j}C_{ij}
\]
where $C_{ij}$ is the correlation function obtained from the data. Because
of the symmetry of the covariance matrix we can use Cholesky factorization
to invert {\bf M}. For the signal autocorrelation function, we used a flat spatial
temperature fluctuation spectrum parametrized by the quadrupole amplitude
$\Delta T$ (see e. g., \cite{Schaefer95})
\begin{equation}
   C_\ell = {6 \sqrt(\pi) \over 5}  (\Delta T)^2 { 2 \ell + 1 \over \ell (\ell
+1)}.\label{eqn:corr}
\end{equation}
$\Delta T$ is then fit to the data and the result is converted to band
power. We obtain very consistent results separately for channels 1
and 2 which are shown in table 1 along with the combined channel results.
\placetable{tbl-1}
Results for channel 1 and 2 are in Table~{\ref{tbl-1}}. The reported 68\% confidence
levels have been calculated in the usual way (see \cite{church}),
and do not include calibration uncertainty. In Figure
~{\ref{fig3}}, we show two theoretical structure formation models
for comparison, the ``standard'' cold dark model and a spatially
flat cosmological constant ($\Lambda$) dominated model.
We see that our data are consistent with the rise to the "Doppler peak"
expected in currently popular adiabatic theories of structure formation
(like the models shown.)

The cosmic origin of our signal has been further checked with the
following null test: we divided the set of the stacked data files
in two subsets; the likelihood analysis of their sum and difference
produced respectively results consistent with those in
Table~~{\ref{tbl-1}}, and with zero (see \cite{romeo2000}) .

    We can also calculate the overlap window functions between different lock-in
patterns and angular lags and use it to estimate the correlations  between
harmonics, similar to the method in (\cite{netter}). The amplitude of the correlated
harmonics was calculated  using these overlap window functions and a flat CMB
spectrum, normalized so the diagonal elements are unity. We found that
correlations between signals measured by the various harmonics are for the
most part negligible ($<1\%$).    Even-odd pairs of harmonics are completely
uncorrelated.   There is some  non-negligible correlation between pairs of
either odd-odd or even-even harmonics.   The correlations between harmonic
pairs 1-3, 2-4, 3-5, and 4-6 are 0.42, 0.53, 0.61 and 0.67, respectively.
We verified that these correlation terms are detectable in a likelihood analysis
of the data and are consistent with our estimates.
The correlations between pairs 1-5 and
2-6 are much smaller: 0.03 and 0.04, respectively. If one wants to be
conservative in using these points to constrain
cosmological parameters, one can use one even-odd pair of harmonics and
be assured that those two points will be completely independent.

  Finally we compare our results with other experiments in Figure 3.  We
see that our lowest l-band temperature measurement is completely
consistent with COBEs highest l-band temperature fluctuation
measurement.  Furthermore, we plot results from two recent balloon
experiments done at similar wavelengths, BOOMERanG
(\cite{deBernardis}), and MAXIMA (\cite{Hanany}).   (For a more complete
summary of experimental results, see Max Tegmark's CMB web site at
http://www.sns.ias.edu/max/cmb/experiments.html.) We see that our results
are remarkably consistent with those of other
experiments, which observed a different part of the sky, and used
completely different experimental and analysis techniques.

\placefigure{fig3}

\section{Conclusion}
   Improvements in our instrumental setup over our 1994 campaign, as
well as refinements in our data analysis procedure, have allowed us to
detect temperature fluctuations over a wide range of angular scales.  We
have used data from our 4-band detector to reduce the atmospheric noise
contamination and to delineate regions of our scans for which the
galactic contamination is insignificant.   We find positive detections
of residual temperature fluctuations in 6 angular ranges in our two
longest wavelength bands (2.1 mm and 3.1 mm).
These measurements, summarized in table 1, are our main result.   The
derived temperatures  are consistent with those seen in other
experiments . Our results essentially show a monotonically increasing
amplitude of temperature fluctuation with l, and are consistent with
some currently popular adiabatic theories of cosmological structure
formation.

\acknowledgments
This work was supported by a University of
Delaware Research Foundation (UDRF) grant, and by the Bartol
Research Institute. We also want to thank the technical staff of
the Observatorio del Teide, and finally R. Hoyland, B. Watson,  L.
Shulman and G. Poirer for their help.

\clearpage

\figcaption[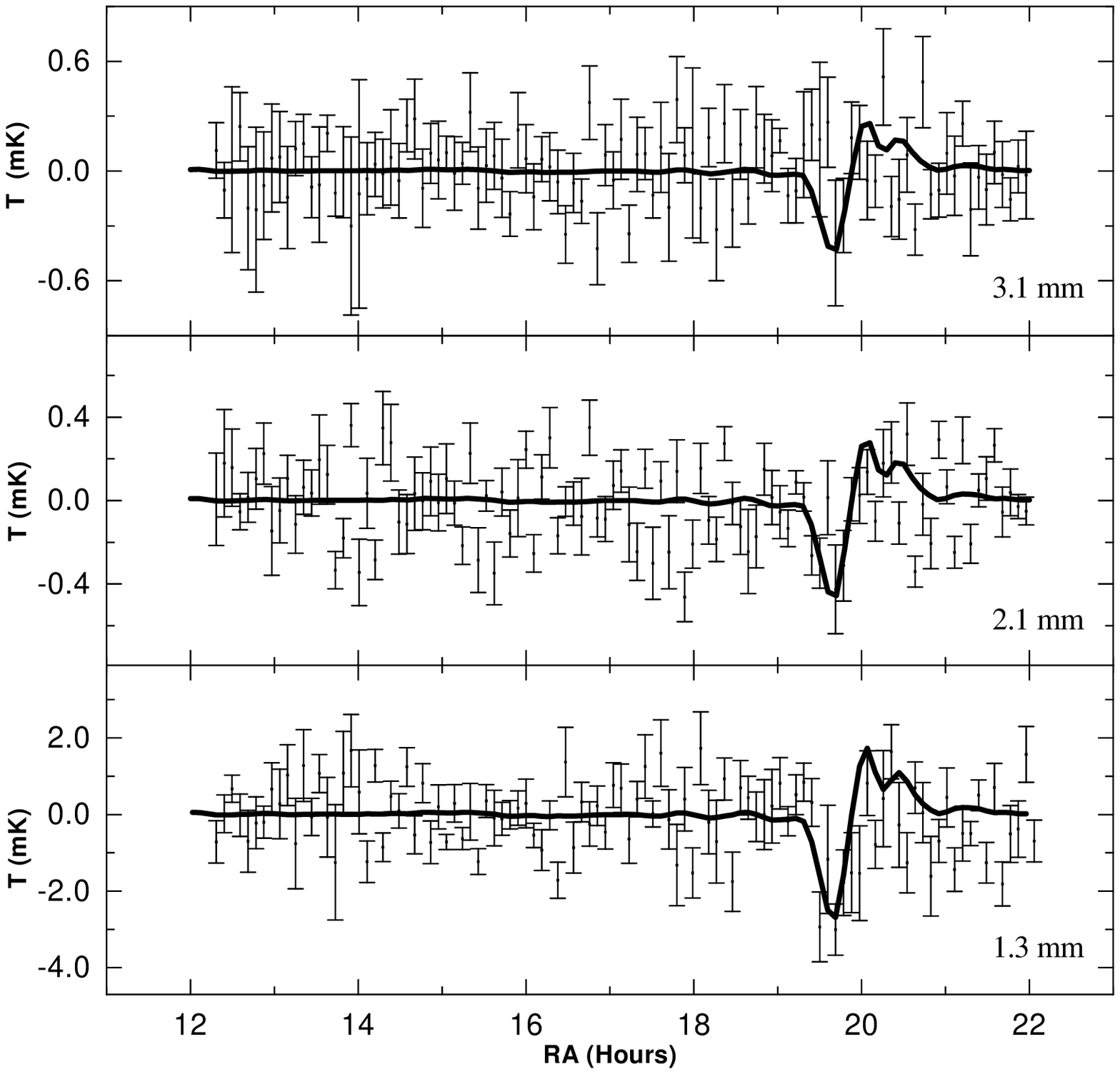]{The first harmonics of  thermodynamic
temperature differences in the three channels.The thick line is
the result of simulating our observing pattern on the
DIRBE 240 $\mu$ m map of the sky, scaled to the corresponding wavelength.
The scaling factors are respectively 5.0, 5.4, and 30.8 $\mu$K/MJy/sr for our
3.1 mm, 2.1mm, and 1.3 mm wavelength channels.\label{fig1}}

\figcaption[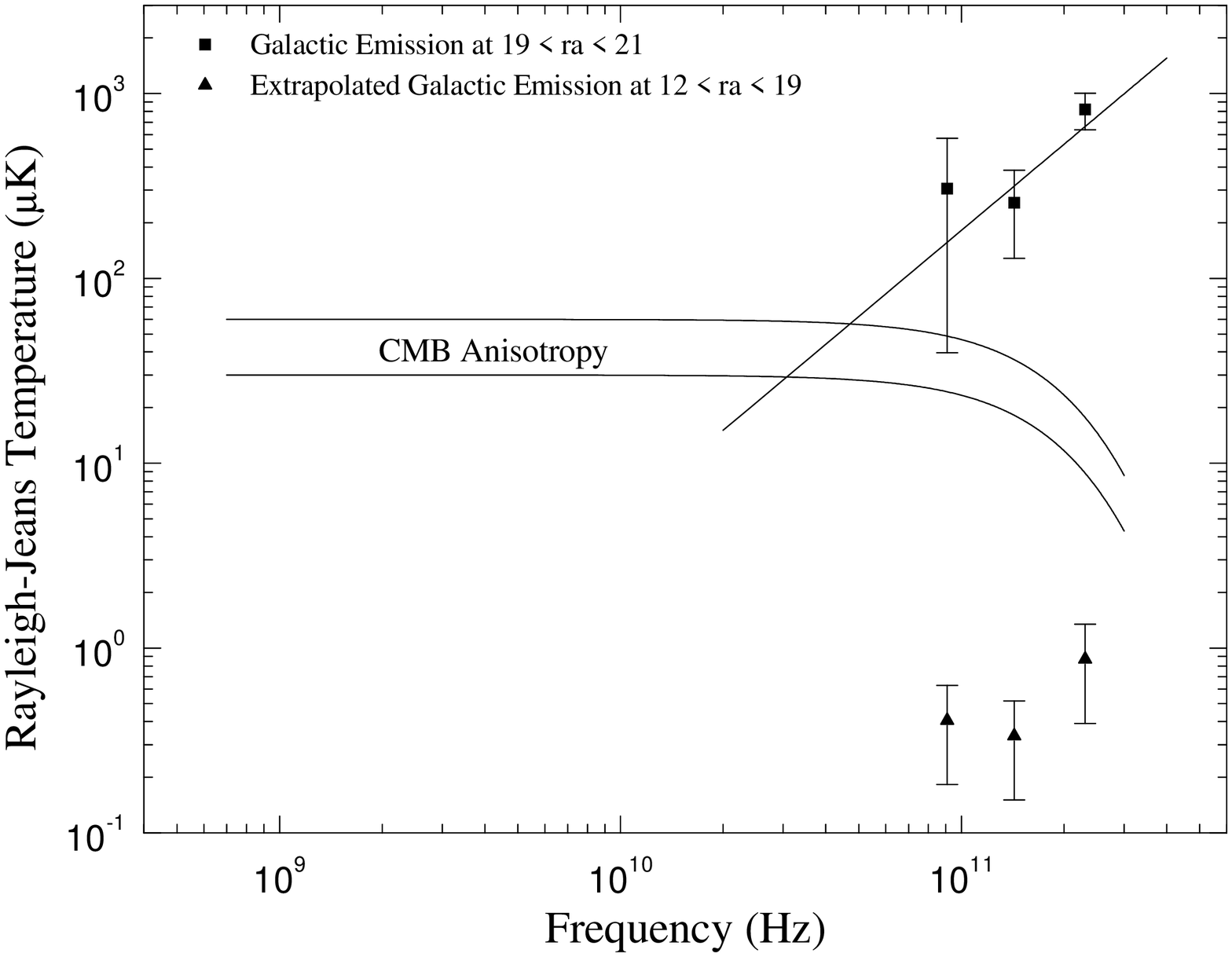]{%
Comparison between Galactic Dust Emission and CMB anisotropy.
The galactic dust emission for each channel has been evaluated in the
galactic plane crossing  region (RA = 19-21 hours) and extrapolated to
the other region (RA = 12 - 19 hours), using information from the first
harmonic on real data and the simulated observations on the DIRBE map
(see figure 1).  This is done for channels 1, 2, and 3.  The spectral
index of the galactic dust emission is estimated to be  $1.5 \pm 1.$ from
fitting a power law line to the data shown above.\label{fig2}}

 \figcaption[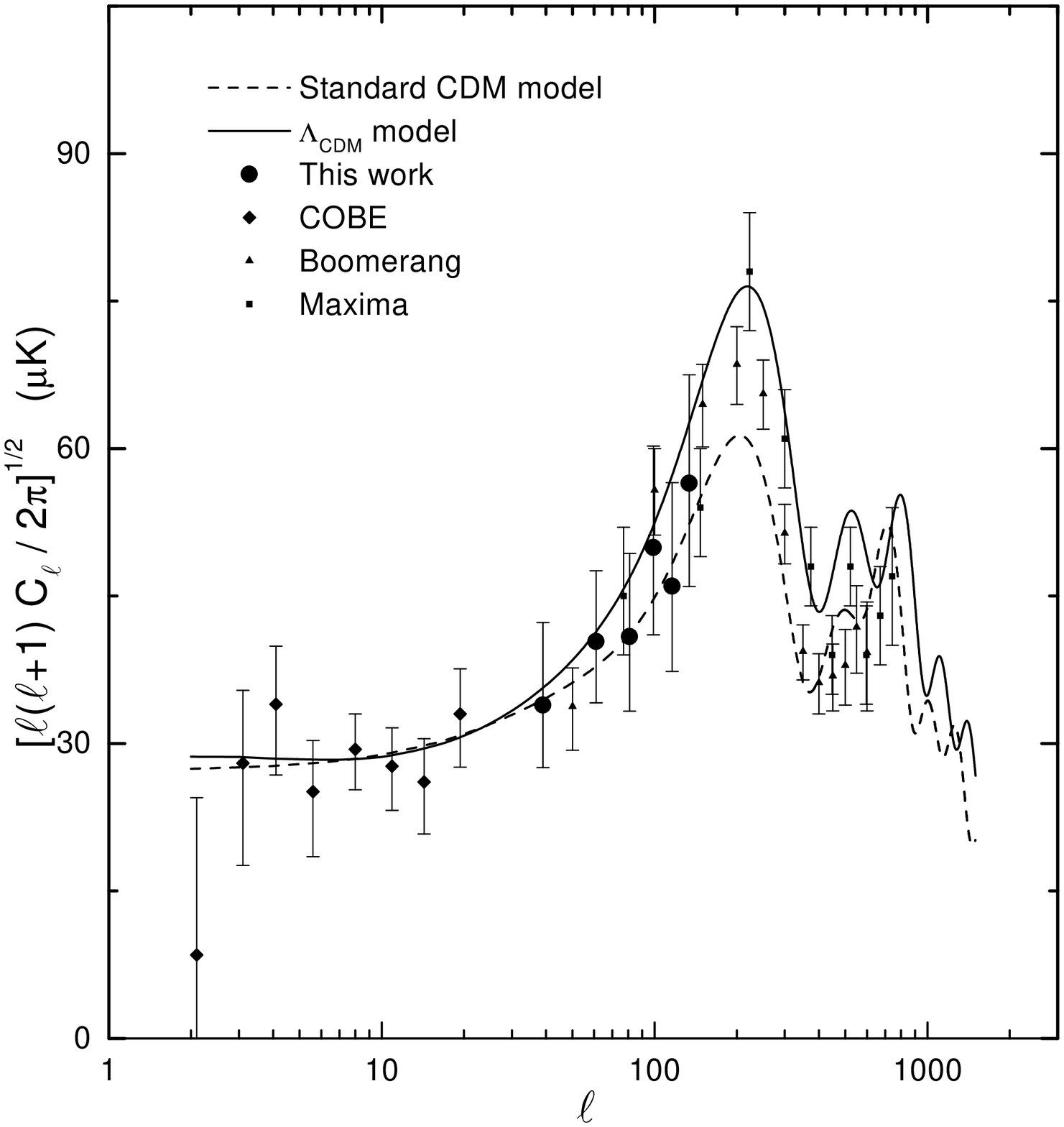]{Band power vs. l for all harmonics.
 We have obtained the theoretical
curves using CMBFAST (\cite{cfast}). Shown are "standard
CDM" ($\Omega_{CDM}$, $\Omega_\Lambda$, $\Omega_b$, $n$, $h = H_0/100$ $km$
$s^{-1}$ $Mpc^{-1}$) = (0.95, 0.00, 0.05, 1.0, 0.65) and $\Lambda$ CDM (0.35, 0.60,
0.05, 1.0, 0.65).  The data shown are from this experiment, COBE (\cite{tegmark}),
and two other recent experiments using similar
wavelengths - BOOMERanG (\cite{deBernardis}) and MAXIMA (\cite{Hanany}).\label{fig3}}

\clearpage

\begin{deluxetable}{cccccccccccc}
\tablecaption{Summary of Band~Powers $\mu$K for channel 1 and 2 as
determined by the likelihood analysis. \label{tbl-1}}
\tablewidth{0pt} \tablehead{ \colhead{Harmonic} &
\colhead{$\bar{\ell }$} &\colhead{Channel 1} &\colhead{Channel
2} & \colhead{Channel 1 + Channel 2}} \startdata
1&$39_{-24}^{+38}$&$33_{-9}^{+11}$&$35_{-9}^{+13}$&$34_{-6}^{+8}$\nl
2&$61_{-22}^{+28}$&$41_{-10}^{+12}$&$40_{-8}^{+9}$&$40_{-6}^{+7}$ \nl
3&$81_{-20}^{+27}$&$43_{-14}^{+16}$&$40_{-9}^{+10}$&$41_{-8}^{+8}$ \nl
4&$99_{-18}^{+24}$&$48_{-15}^{+17}$&$51_{-11}^{+13}$&$50_{-9}^{+10}$ \nl
5&$116_{-14}^{+23}$&$49_{-14}^{+18}$&$44_{-11}^{+13}$&$46_{-9}^{+10}$ \nl
6&$134_{-12}^{+20}$&$60_{-35}^{+28}$&$56_{-11}^{+12}$&$56_{-10}^{+11}$\nl
\enddata
\end{deluxetable}

\clearpage

\plotone{note1.eps} \clearpage
%\epsscale{0.5}
\plotone{galactic_cont2.eps}\newline
\plotone{results.eps} %\clearpage

\end{document}